# On Properties of Phase-Conjugation Focusing for Large Intelligent Surface Applications – Part II: Horizontal Polarization

Jiawang Li[1]
[1] Department of Electrical and Information Technology, Lund University, 221 00 Lund, Sweden
jiawang.li@eit.lth.se

***Abstract*— Near-field focusing (NFF) forms the basis for several applications of large intelligent surface (LIS) in sub-10 GHz bands, including wireless communications, wireless power transfer, positioning, and sensing. In this two-part paper, Part I analyzed the properties of phase conjugation NFF for vertically polarized antennas, in a circular array configuration. In Part II of this article, we continue to study phase conjugation NFF for circular arrays, but for horizontally polarized antennas. We investigate the focusing characteristics of a circular array in two distinct configurations. The numerical results show that the first configuration where all the antenna elements (including the user antenna) are aligned offers significant better performance in terms of peak gain, 3 dB focal width and sidelobe level, relative to the second configuration where the broadside of the elements faces the array center. This result points to the beneficial use of orthogonally oriented horizontally polarized antenna at the user to allow for flexible polarization alignment with the fixed array orientation. In addition, the vertical polarized circular array of Part I may be merged with the first configuration of Part II to provide optimal NFF to a randomly oriented user equipment with a polarization reconfigurable tripole antenna.**

## I. Introduction

As mentioned in Part I, large intelligent surface (LIS) [1] is an innovative wireless system concept poised to meet the demanding requirements of 6G in sub-10 GHz frequency bands, despite the constraints of limited spectrum resources. Unlike phased array systems, which are typically analyzed in the far-field (FF) [2]-[3], since the user equipment (UE) of LIS operates in the near-field (NF) region, the transmission differs from the plane waves typically observed in the FF region. The impact of spherical wave propagation in the near field must be carefully evaluated [4]. Whether the LIS system is designed for information [5] or power [6] transmission, or some other applications, it is essential to focus the power to boost the signal-to-noise ratio, transmission efficiency and/or spatial resolution.

In this context, this two-part paper takes up the task of exploring how LIS can be configured to yield desirable NF focusing (NFF) properties, especially since previous work on NFF does not consider the scenario where the user is surrounded by LIS antenna elements from more than one side. As in Part I, phase conjugation [7] is utilized in this part (Part II) to perform NFF, due to its simplicity and convenient physical interpretation. Moreover, a uniform circular array is used as a simple LIS array, with the user equipment (and its antenna) located inside the circular array, and on the same two-dimensional (2D) plane as the array. However, the focus here is on horizontal polarization, and as opposed to the relatively simpler setup of vertical polarization in Part I, horizontal polarization is nonunique in the NF, given the possibility of arbitrary horizontal orientation of the array elements.

The NFF insights obtained from the study into horizontal polarization in Part II complement the findings of Part I relating to vertical polarization, allowing 2D LIS array to be configured to yield desirable NFF properties for different LIS applications.

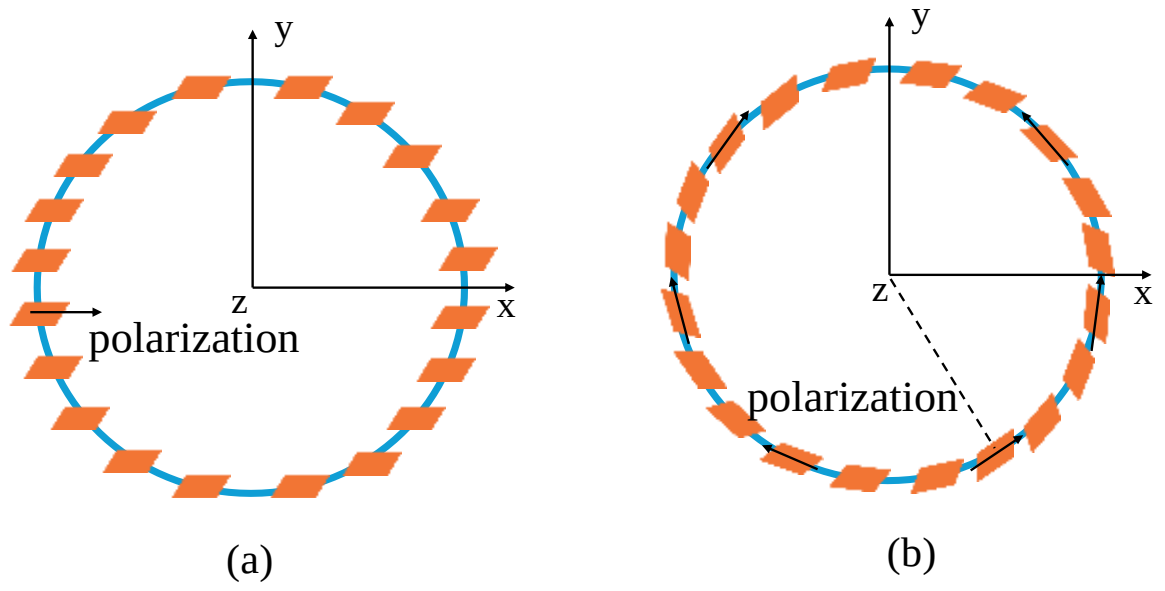

Fig. 1. Two configurations of UCAs: (a) all elements aligned with x-axis (C1), (b) element broadside facing array center (C2).

## II. NFF Properties of Circular Arrays

As depicted in Fig. 1, two configurations of horizontally polarized uniform circular arrays (UCAs) are selected in this study. In the first configuration (C1), the polarization direction is aligned with the *x*-axis for all array elements, whereas the UE antenna is *x*- or *y*-polarized. In the second configuration (C2), the broadside of each element faces the array center, and the UE's single antenna is *x*-polarized. The two configurations are motivated by a fixed polarization alignment (C1) and a circularly symmetric alignment (C2). C1's UE antennas are polarized to represent extreme cases (co-/cross-polarization). Moreover, due to the circular symmetry in C2, arbitrary horizontal orientation of the UE antenna will result in similar NFF properties.

As in Part I, several simplifying assumptions are made for the array elements in the numerical study: negligible mutual coupling, identical radiation patterns in the 2D plane, and single horizontal (linear) polarization. These assumptions can be fair approximations of real cases, e.g., horizontal dipole elements with over ½ wavelength ($\lambda$) spacing between the closest pair. It is further assumed that there is no scattering by either object in the surroundings or the other array elements.

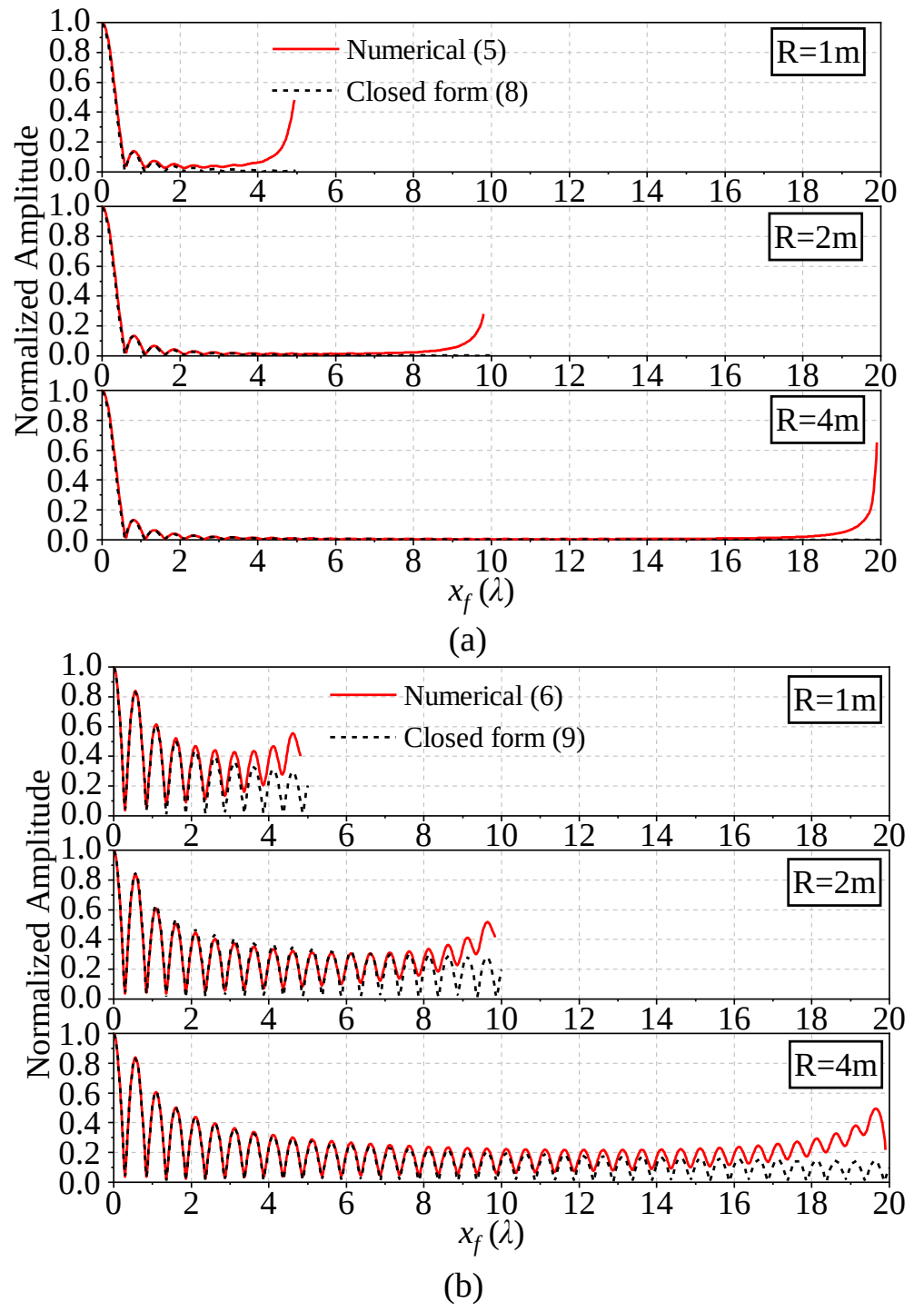

Fig. 2. Numerical and closed form values for normalized (a) $E_x$ and (b) $E_y$ for different radii with $N$ = 120 and $\lambda$ = 0.2 m.

To demonstrate the impact of horizontal polarization more explicitly, here we further assume that the array and UE antennas are Hertzian dipoles, based on which the electric FF vector components may be derived.

In the FF region and equally applicable to radiative NF region (i.e., spherical wave propagation), the vector potential $\boldsymbol{A}$ of a Hertzian dipole (centered at the origin) at the observation point $\boldsymbol{r} = r\boldsymbol{a}_r$ is [8]

$$\boldsymbol{A}(\boldsymbol{r}) = \boldsymbol{a}_l \frac{I_0 \mu_0 l}{4\pi r} e^{-jkr}, \qquad (1)$$

where $r$ is the distance from the dipole to the field point, $l$ is the infinitesimal dipole's length, $\boldsymbol{a}_l$ is the unit vector in the direction of the current $I_0$, and $\mu_0$ is the permeability in vacuum. The electric field is [8]

$$\boldsymbol{E} = -j\omega\boldsymbol{A} - \nabla\phi == -j\omega\boldsymbol{A} - j\frac{1}{\omega\mu_0\varepsilon_0}\nabla(\nabla\cdot\boldsymbol{A}), \qquad (2)$$

where $\phi$ is the scalar potential and $\varepsilon_0$ the dielectric constant in vacuum. Considering the polarization of the dipole, and assuming $I_0$ is a constant

$$\boldsymbol{E}(\boldsymbol{r},\boldsymbol{l}) = \frac{j\eta I_0 lk}{4\pi r} e^{-jkr}\left(\boldsymbol{a}_r \times (\boldsymbol{a}_l \times \boldsymbol{a}_r)\right), \qquad (3)$$

where $\eta$ is the wave impedance in free space.

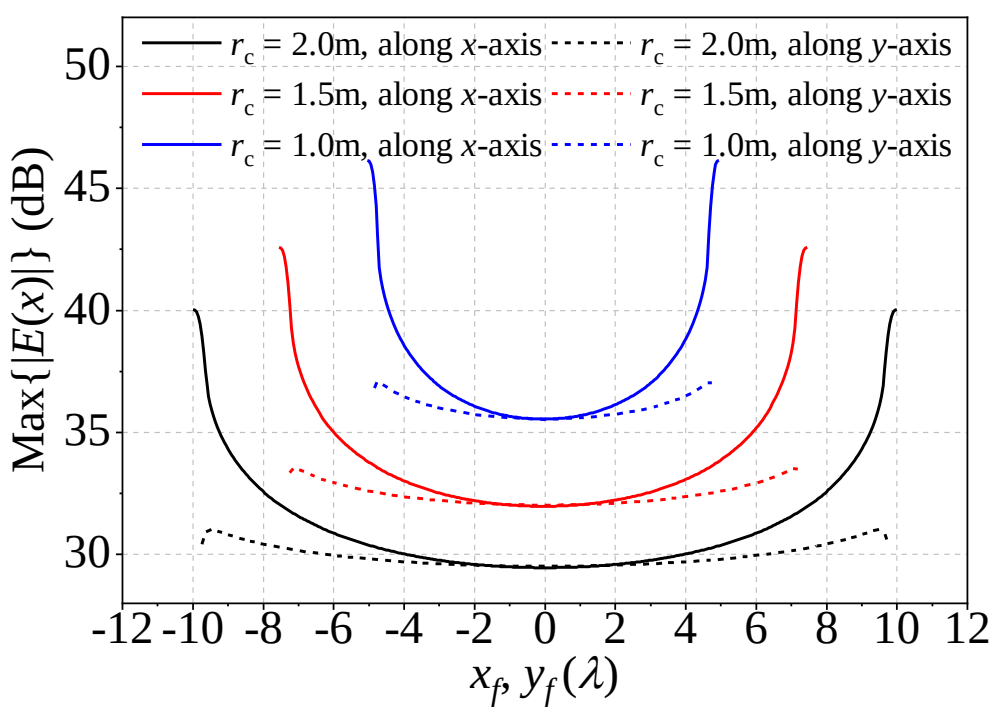

Fig. 3. Peak gain for configuration C1 with focal point along the $x$- and $y$-axes, for $x$-oriented receiving Hertzian dipole.

For an $N$-element circular array with radius $r_c$, the source (dipole) point for element $n$ is $\boldsymbol{r}_n$. Therefore, $r = |\boldsymbol{r} - \boldsymbol{r}_n|$ in (3) and the current direction in dipole $n$ is $\boldsymbol{a}_{ln}$. The total electric field as received by at $\boldsymbol{r}$ is then

$$\boldsymbol{E}(\boldsymbol{r},\boldsymbol{l}) = \sum_{i=1}^{N} \frac{j\eta I_0 l_n k}{4\pi|\boldsymbol{r} - \boldsymbol{r}_n|} e^{-jkr_n}\left[\boldsymbol{a}_{rn} \times (\boldsymbol{a}_{ln} \times \boldsymbol{a}_{rn})\right]. \qquad (4)$$

As pointed out in Part I, the waves from different array elements arrive at the receiving antenna from different directions, unlike the FF case. However, the presence of antennas, with (4) implicitly assuming an $x$-polarized and a $y$-polarized Hertzian dipole receiving the corresponding $x$ and $y$-electric field components, allows the wave propagation directions to be neglected. To proceed, $\boldsymbol{a}_{rn} \times (\boldsymbol{a}_{ln} \times \boldsymbol{a}_{rn}) = \boldsymbol{a}_{ln} - (\boldsymbol{a}_{rn} \bullet \boldsymbol{a}_{ln})\boldsymbol{a}_{rn}$ and using Cartesian coordinates, $\boldsymbol{r} = (x, y)$, $\boldsymbol{r}_n = (x_n, y_n)$, $x_n = r_c\cos\theta_n$, and $y_n = r_c\sin\theta_n$. If $N$ is even, $\theta_n = \frac{2\pi(n+1)}{N}$ and if $N$ is odd, $\theta_n = \frac{2\pi n}{N}$.

## A. *Configuration C1*

Since all the array elements in this configuration are $x$-polarized, $\boldsymbol{a}_{ln}$ = (1, 0). The $x$- and $y$-components of the total electric field (4) can be obtained as follows

$$E_x(x,y) = \frac{j\eta I_0 lk}{4\pi}\sum_{n=1}^{N}(y - \sin\theta_n)^2 P(x,y), \qquad (5)$$

$$E_y(x,y) = \frac{j\eta I_0 lk}{4\pi}\sum_{n=1}^{N}(x - \cos\theta_n)(y - \sin\theta_n)P(x,y), \qquad (6)$$

where

$$P(x,y) = \frac{e^{-jk\sqrt{(x - r_c\cos\theta_n)^2 + (y - r_c\sin\theta_n)^2}}}{\left[(x - r_c\cos\theta_n)^2 + (y - r_c\sin\theta_n)^2\right]^{3/2}}. \qquad (7)$$

When analyzing the properties of the circular array, the phase conjugation method ensures that the phase term in the numerator of (7) is cancelled out. Furthermore, amplitude normalization is performed such that $\eta I_0 lk/(4\pi) = 1$. Here, the electric field distribution for the focal point at (0, 0) is explored in more detail. First, the amplitudes of $E_x$ and $E_y$ normalized to the values at (0, 0) (i.e., $F_x$ and $F_y$) are calculated from (5) and (6) for positive $x$-axis (see Fig. 2). Moreover, for large $N$, closed forms can be derived (see Appendix A) as

$$F_x(r) \approx \begin{cases} \left| \dfrac{J_1(2\pi r_c)}{\pi r} \right|, & r \neq 0 \\ 1, & r = 0 \end{cases}, \tag{8}$$

$$F_y(r) \approx \left| J_0(2\pi r) - J_2(2\pi r) \right|. \tag{9}$$

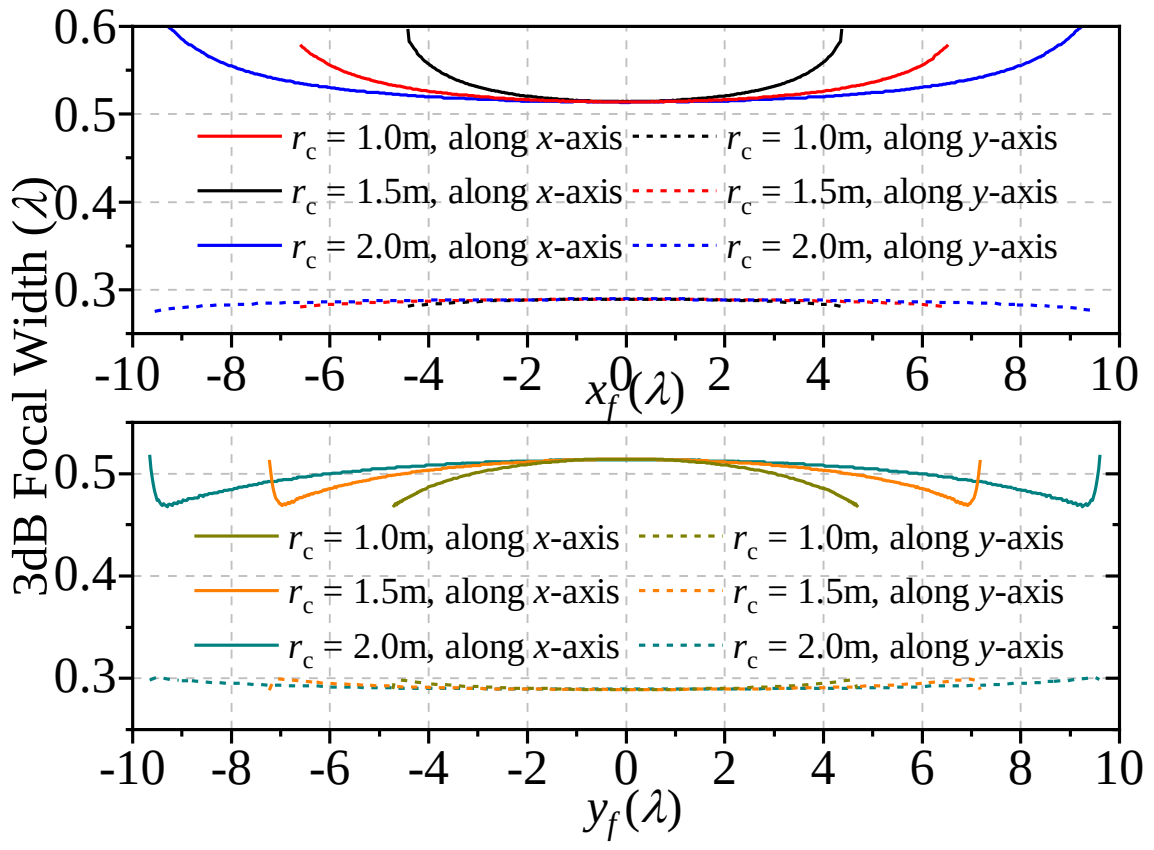

Fig. 4. 3 dB focal width (along $x$- and $y$-axis) of $E_x$ in C1 as the focal point moves along the $x$- (top subplot) and $y$-axes (bottom subplot).

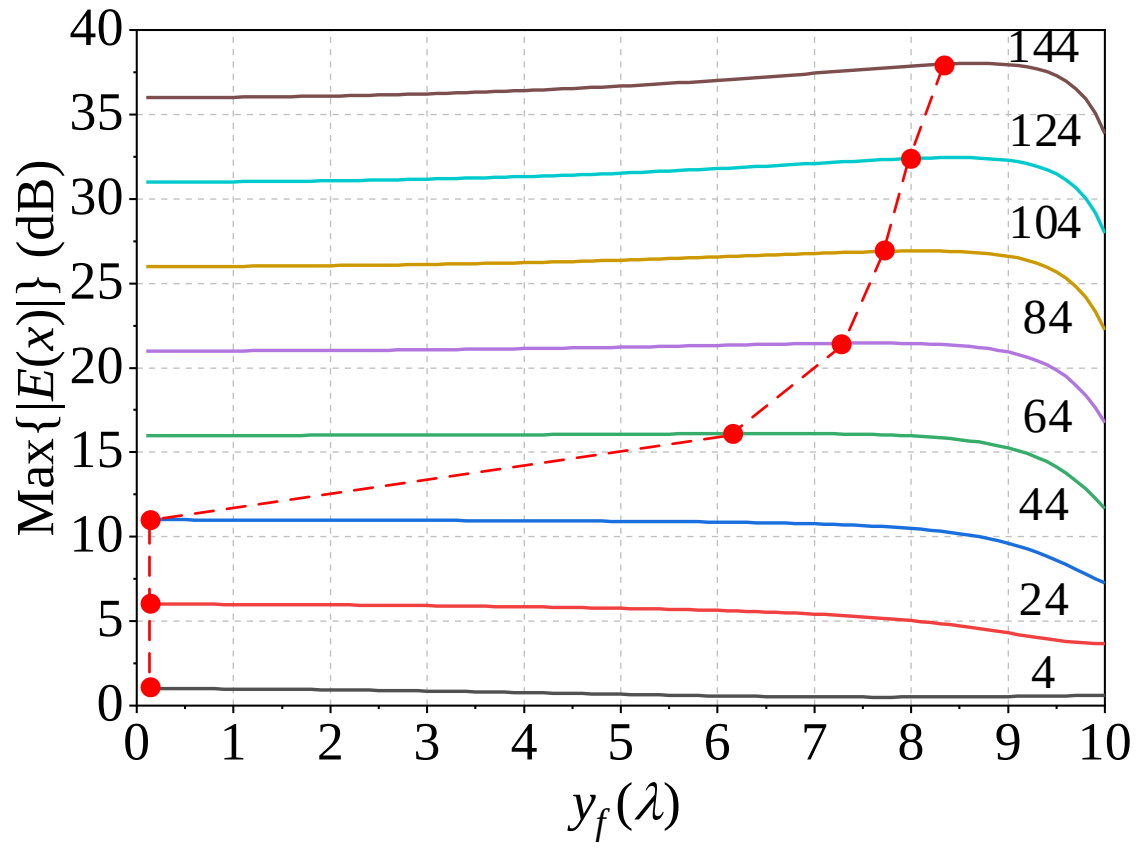

Fig. 5. Peak gain of C1 for focal point along the $y$-axis, for $x$-oriented receiving Hertzian dipole. The number next to each curve indicates the number of elements $N$.

The approximate closed form values in (8) and (9) are also shown in Fig. 2 for comparison. The closed form results agree well with the numerical results, except for the region close to the array edge. Furthermore, $F_y$ has higher sidelobes than $F_x$, which is undesirable for NFF. In addition, as will be presented later, the $E_y$ is significantly lower than $E_x$ in its peak gain, which highlights the effect of polarization mismatch between the array and UE antennas.

*1) NFF Properties of $E_x$*

Figure 3 shows the results of the peak gain for $E_x$ with $N$ = 120, $\lambda$ = 0.2 m and array radius of $r_c$ =1 m, 1.5 m, and 2 m. As $r_c$ increases, the gain decreases, especially when $r_c$ goes from 1 m to 2 m (i.e., a 3 dB drop at $x_f$ = 0). This 3 dB drop can be derived for a sufficiently large $N$, where the summation in (5) can be well approximated by an integral, allowing for the proportionality relationship $E_x(0,0) \propto \pi / r_c$ to be derived for the focal point at (0, 0). The 3 dB focal widths as the focal point moves along the $x$- and $y$-axes are depicted in Fig. 4. The focal width is relatively stable inside the array. Within ±2$\lambda$ from the center, it is approximately 0.518λ along the $x$-axis and around 0.29$\lambda$ along the $y$-axis (for both $x_f$ and $y_f$ sweeps).

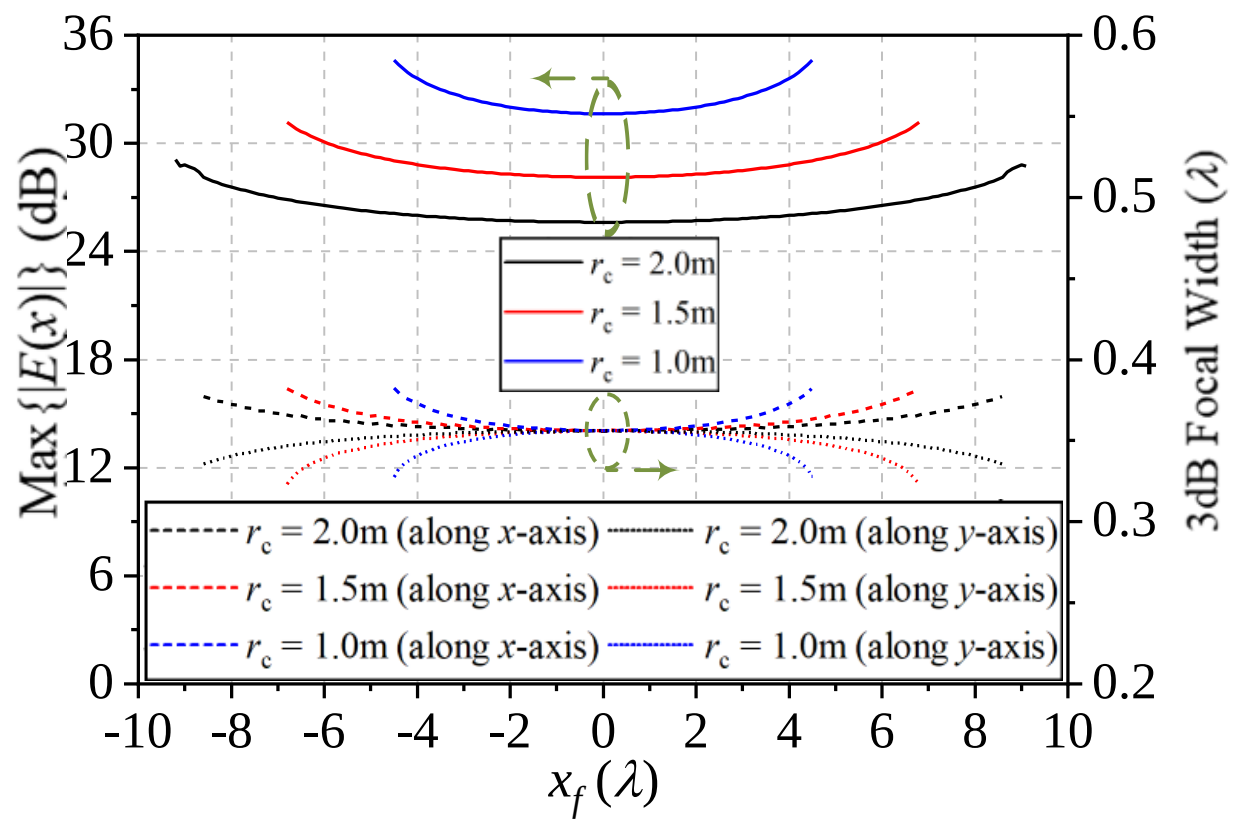

Fig. 6. Peak gain and focal width for C1 with focal point along the $x$-axis, for $y$-polarized receiving Hertzian dipole.

It is observed in Fig. 3 that the gain fluctuation within the array is less than 1 dB for focal point along the $y$-axis, for $N$ = 120 and $r_c$ = 2 m, and the maximum peak gain occurs close to the array edge. To study the peak gain behavior further for focal point along the y-axis, it is plotted in Fig. 5 for different $N$'s. As illustrated in Fig. 5, when $N$ is small, the maximum peak gain is located at the center of the array. But as the array size increases, the maximum peak gain progressively shifts toward the edge of the array.

*2) NFF Properties of $E_y$*

The properties of the $E_y$ electric field are presented in Fig. 6, showing that the trends in peak gain and focal width are similar to those of $E_x$. Due to symmetry in the electric field magnitude distribution along $x$ and $y$-axes (see Fig. 7, as will be elaborated later), only $x_f$ needs to be analyzed. However, differences are observed in the absolute values of peak gain and focal width. These differences are primarily attributed to the $\sin^2(\bullet)$ term in $E_y$. Due to the circular symmetry, there is a 180° phase difference between elements concerning the center. The conjugate-phase method accounts for the combined phase differences from both the path and the element structure, So the result should be as follows:

$$E_y(\Delta,0) = \int_0^{2\pi} \frac{r_c \left|\Delta - r_c\cos\theta\right| \left|\sin\theta\right| e^{-jk\left(\sqrt{(\Delta - r_c\cos\theta)^2 + (r_c\sin\theta)^2} - r_c\right)}}{\sqrt{(\Delta - r_c\cos\theta)^2 + (r_c\sin\theta)^2}^{\,3}} \tag{10}$$

Similar with the Appendix A, the formula (14) can be simplified as:

$$E_y(\Delta,0) = \int_0^{2\pi} \frac{\left|\cos\theta\right| \left|\sin\theta\right| e^{-jk\left(\sqrt{(\Delta - r_c\cos\theta)^2 + (r_c\sin\theta)^2} - r_c\right)}}{r_c} d\theta \tag{11}$$

Then it can get the approximate solution as follows (the maximum value is $1/r_c$ at (0, 0)),

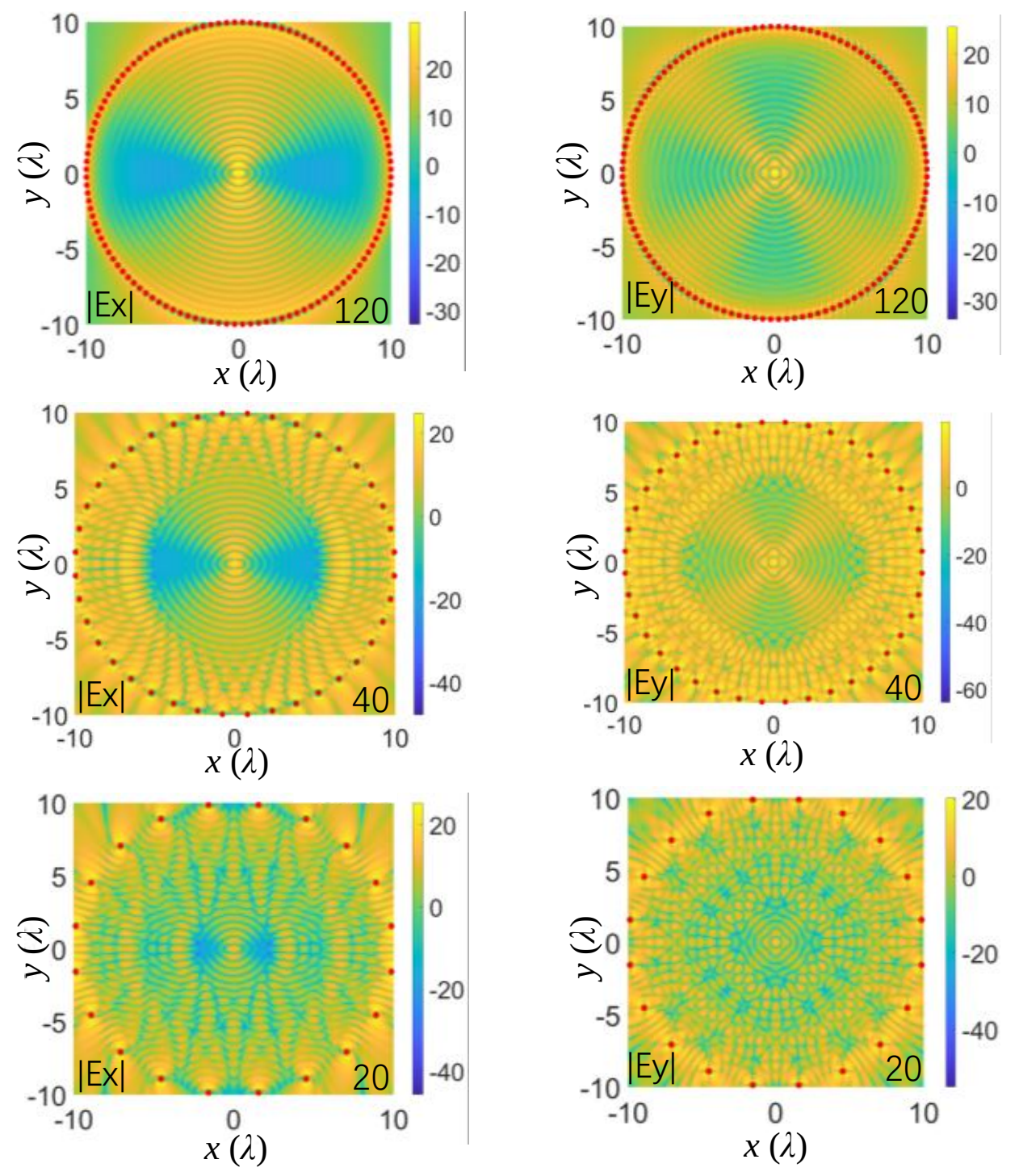

Fig. 7. Normalized electric field distribution across different numbers of antenna elements for C1.

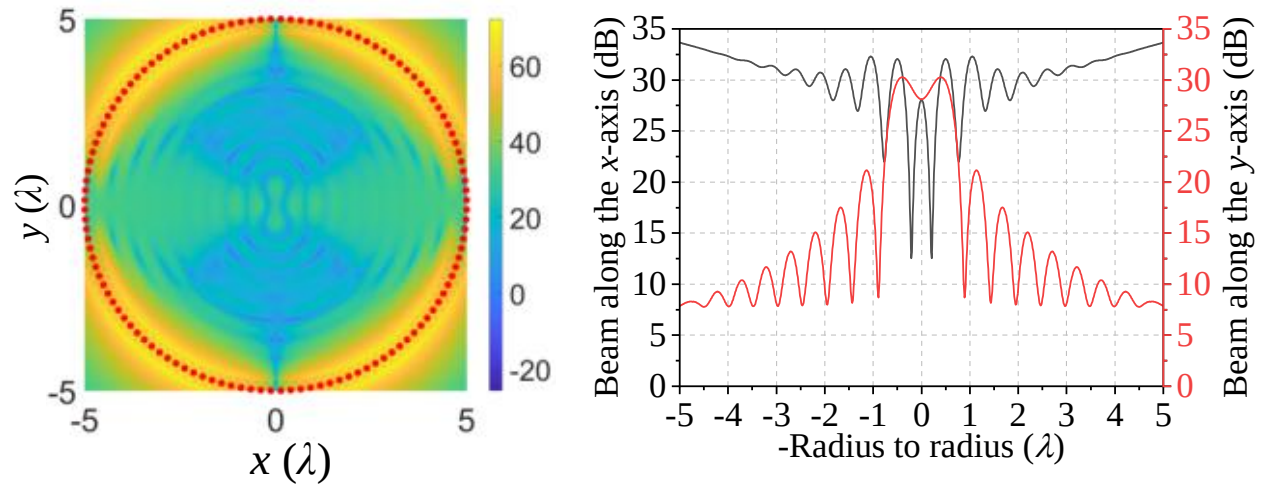

(a) (b)

Fig. 8. Electric field distribution for $N$ = 60 of C2 with the radius of 1 m (a) in 2D plot and (b) along the *x- and y*-axes.

$$E_y(\Delta,0)=\begin{cases}\dfrac{\lambda\left|\lambda-\lambda\cos\left(\dfrac{2\pi\Delta}{\lambda}\right)-2\pi\Delta\sin\left(\dfrac{2\pi\Delta}{\lambda}\right)\right|}{\pi^2\Delta^2 r_c}, & \Delta\neq 0\\ 2/r_c, & \Delta=0\end{cases} \quad (12)$$

From (12), the 3 dB focal width at the array center is ~0.36$\lambda$. Its beamforming gain is 3.92 dB lower than that of $E_x$.

When $N$ is large enough, the summation in (6) can be approximated by an integral, resulting in $E_y(0,0)=2/r_c$ if $\eta I_0 lk/(4\pi)=1$ is assumed. This together with the previous result of $E_x(0,0)\propto\pi/r_c$ are interesting in the sense that an increase in the number of antenna elements will increase the array gain proportionally in FF beamforming. However, for horizontally polarized dipoles, the behavior differs in NFF, where the peak gain remains constant at the center of the array when the entire circular array radiates power.

Figure 7 shows how the focal width as well as the grating lobes and sidelobes change for both $|E_x|$ and $|E_y|$ when the focus is at (0,0) and $N$ = 20, 40, and 120, respectively. The red points in the figures represent the antenna elements. Due to the structure and polarization of the array elements, $|E_x|$ is not circularly symmetric around the array. The focusing effect is stronger along the *x*-axis, where lower sidelobes are observed. In contrast, high sidelobes are observed along the *y*-axis. Like vertical polarization, the 3 dB focal region remains unchanged regardless of $N$. $E_y$ exhibits a higher degree of symmetry and generates regions of stronger electric field along the ±45° directions. Its focal width is approximately the same as that of the vertically polarized array in Part I. However, the spreading of energy (i.e., the presence of regions with high sidelobes) is the primary factor contributing to the reduction in the peak gain of the focal region.

### *B. NFF Properties for Configuration C2*

Following the same rules of (4), for C2, the general expression of $|E_x|$ and $|E_y|$ can be obtained as in (10) and (11). Figure 8 shows the electric field within the array aperture when the focal point is located at (0, 0). In addition to a peak gain generated by phase conjugation focusing along the *x*-axis, sidelobes that are even higher than the main lobe are observed. On the other hand, along the *y*-axis, the focusing forms a relatively wide beam. It is worth noting that its peak gain is reduced by ~7.4 dB compared to the C1 configuration, which means far less effective NFF. Due to the poor NFF gain performance of C2, it is not considered practical for real applications.

$$E_x(x,y)=\frac{jP(x,y)\eta I_0 Lk}{4\pi}\sum_{n=1}^{N}\begin{pmatrix} r_c^{\,2}\begin{pmatrix}\sin\theta_n\,\mathrm{sgn}(\cos\theta_n)-\cos\theta_n\sin^2\theta_n\,\mathrm{sgn}(\cos\theta_n)\\ -|\cos\theta_n|\sin^2\theta_n\end{pmatrix}\\ +x^2\left(\sin\theta_n\,\mathrm{sgn}(\cos\theta_n)\right)+y^2\begin{pmatrix}\mathrm{sgn}(\cos\theta_n)\sin\theta_n-\\ |\cos\theta_n|\end{pmatrix}\\ +xr_c\left(\sin^2\theta_n\,\mathrm{sgn}(\cos\theta_n)-2\cos\theta_n\sin\theta_n\,\mathrm{sgn}(\cos\theta_n)\right)\\ yr_c\begin{pmatrix}2|\cos\theta_n|\sin\theta_n+\cos\theta_n\sin\theta_n\,\mathrm{sgn}(\cos\theta_n)-\\ 2\sin^2\theta_n\,\mathrm{sgn}(\cos\theta_n)\end{pmatrix}-\\ xy\,\mathrm{sgn}(\cos\theta_n)\sin\theta_n\end{pmatrix} \quad (13)$$

$$E_y(x,y)=\frac{jP(x,y)\eta I_0 Lk}{4\pi}\sum_{n=1}^{N}\begin{pmatrix} r_c^{\,2}\begin{pmatrix}|\cos\theta_n|-\cos\theta_n\sin^2\theta_n\\ \mathrm{sgn}(\cos\theta_n)-|\cos\theta_n|\sin^2\theta_n\end{pmatrix}-x^2|\cos\theta_n|\\ +xr_c\begin{pmatrix}\mathrm{sgn}(\cos\theta_n)\sin^2\theta_n\\ -2|\cos\theta_n|\cos\theta_n\end{pmatrix}+yr_c\,\mathrm{sgn}(\cos\theta_n)\\ \cos\theta_n\sin\theta_n-xy\,\mathrm{sgn}(\cos\theta_n)\sin\theta_n\end{pmatrix} \quad (14)$$

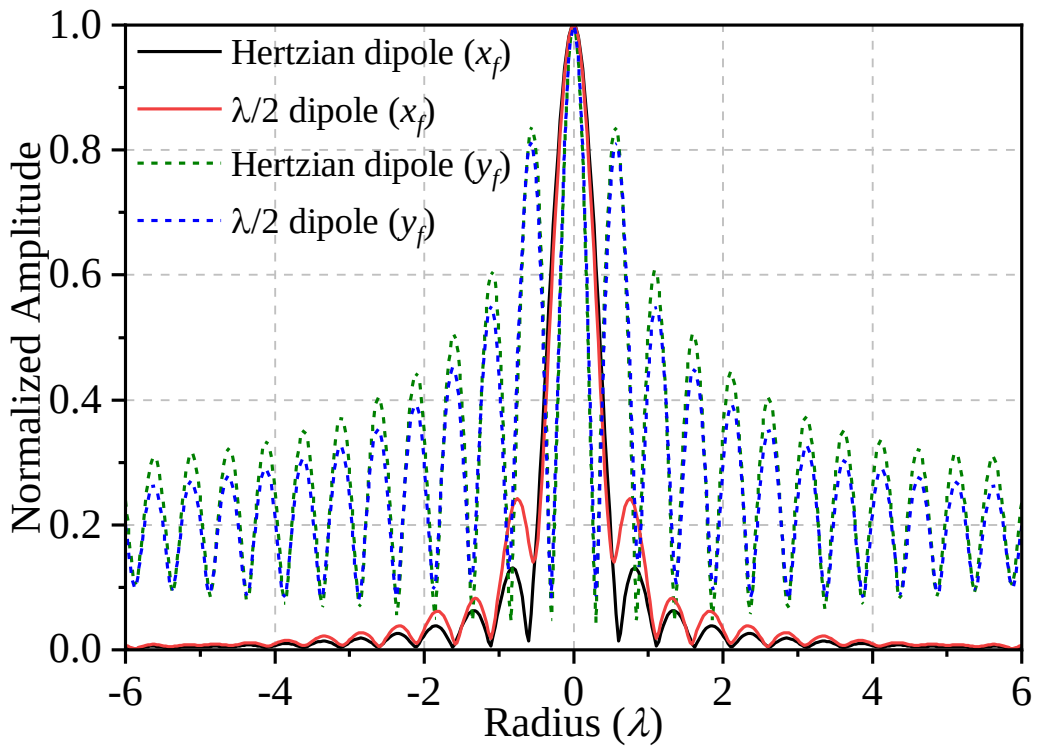

Fig. 9. $E_x$ simulated result in FEKO for Hertzian dipole and half-wavelength dipole when the focal point is at (0, 0).

where

$$\operatorname{sgn}(\cos\theta_n) = \begin{cases} 1, \cos\theta_n \geq 0 \\ -1, \cos\theta_n < 0 \end{cases}. \tag{15}$$

### C. Electromagnetic simulation verification

To verify the accuracy of the proposed model, electromagnetic simulation software (FEKO 2022) is employed to validate the *x*-polarized electric field. The analysis is extended from a Hertzian dipole to a half-wave dipole to enhance practical applicability. The half-wave dipole is modeled as a cylindrical conductor with a length of 100 mm and a radius of 0.1 mm, excited by a port placed at the center of the cylinder. The simulation results are shown in Fig. 9. As expected, the results for the Hertzian dipole agree well with the theoretical predictions. Although the half-wave dipole exhibits some deviations, both types of dipoles produce the same main lobe beamwidth. Due to the different current distribution of the half-wave dipole, the side lobe level of the *x*-directed electric field ($E_x$) slightly increases, while the side lobe level in the y-direction slightly decreases. These results indicate that the theoretical model remains valid and applicable even when a half-wave dipole is used in practical scenarios.

The spacing between antenna elements is also a critical factor influencing transmission efficiency. Initially, as the number of elements increases, beamforming effects between antennas enhance the overall gain. However, when the number becomes large, mutual coupling between elements begins to dominate, causing a decline in efficiency. This indicates the existence of an optimal spacing point that maximizes performance.

As illustrated in Fig. 10, a half-wavelength dipole with the same polarization as the other elements is placed at the center of the coordinate system. By calculating the transmission coefficient between each surrounding element and the central dipole, the total transmission efficiency can be determined. In this analysis, the total input power of the system is assumed to be 1 W. The simulation is conducted with the elements arranged in a circular ring with a radius of 0.5 m.

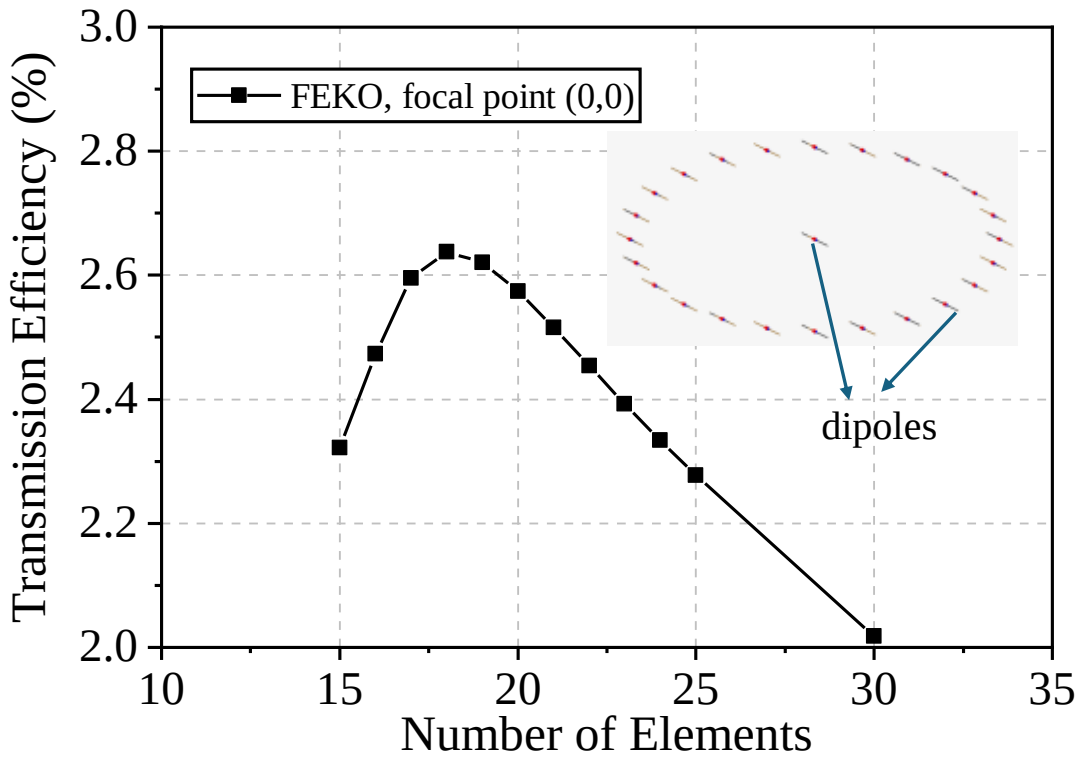

Fig. 10. Transmission efficiency simulated result in FEKO for half-wavelength dipole when the number changes.

The results reveal that the system achieves optimal transmission efficiency when the element spacing is 0.868$\lambda$ and the maximum system efficiency is about 2.64% in this case.

## III. Conclusions

This paper briefly examines the NFF performance of simplified LIS setups using circular arrays for horizontal polarization. For the configuration with all *x*-polarized array elements (C1), it is found that the array can provide stable peak gain and localized focal region $\pm 2\lambda$ from the array center. The $E_x$-field provides higher gain; however, the focal point along *x*- and *y*-axes exhibits significant differences in focal width between the *x*- and *y*-axes. Moreover, for focal point at (0, 0), low sidelobes are observed along the x-axis, whereas high sidelobes are observed along the y-axis sidelobes. In contrast, $E_y$ demonstrates a more symmetrical 2D electric field distribution, but the maximum peak gain is 3.92 dB lower than $E_x$. For C2, where the broadside of array elements faces the array center, the poor NFF performance in terms of peak gain, focal width and sidelobes indicates that it is impractical.

These results point to the beneficial use of orthogonally oriented horizontally polarized antenna at the user to allow for flexible polarization alignment with the fixed array orientation (i.e., aligning the user antenna polarization with the array polarization). In addition, the vertical polarized circular array of Part I may be merged with the first configuration of Part II to provide optimal NFF to a randomly oriented user equipment with a polarization reconfigurable tripole antenna.

Moreover, the C1 configuration exhibits similar NFF trends as the vertical polarized array in Part I, with the peak gain and 3 dB focal width being stable near the array center, with the peak gain increasingly rapidly towards the array edge along the direction of polarization.

## Appendix A

Assuming $r_c \gg \Delta_1$, the electric field along the *x*-axis with $x = \Delta_1$ can be expressed as

$$E_x(\Delta_1,0)=\int_0^{2\pi}\frac{(r_c\sin\theta)^2 e^{-j\frac{2\pi}{\lambda}\left(\sqrt{(\Delta_1-r_c\cos\theta)^2+(r_c\sin\theta)^2}-r_c\right)}}{\sqrt{(\Delta_1-r_c\cos\theta)^2+(r_c\sin\theta)^2}^{3}}. \quad (16)$$

Since $\sqrt{(x-r_c\cos\theta)^2+(r_c\sin\theta)^2}\approx r+\Delta_1^2/2r_c-\Delta_1\cos\theta$ by Taylor's series expansion, substituting it into (16), and according to Euler's formula, and due to $r_c\gg\Delta_1$, $\Delta_1$ in the denominator of (13) can be neglected. Then, the integral in the real and imaginary parts are as follows

$$\mathrm{Re}\left(E_x(\Delta_1,0)\right)=\int_0^{2\pi}\frac{\sin^2\theta\cos\left(\frac{2\pi}{\lambda}\left(\frac{\Delta_1^2}{2r_c}-\Delta_1\cos\theta\right)\right)}{r_c}d\theta \quad (17)$$

$$\mathrm{Im}\left(E_x(\Delta_1,0)\right)=\int_0^{2\pi}\frac{\sin^2\theta\sin\left(\frac{2\pi}{\lambda}\left(\frac{\Delta_1^2}{2r_c}-\Delta_1\cos\theta\right)\right)}{r_c}d\theta \quad (18)$$

After the integration, the final result is

$$E_x(\Delta_1,0)\approx\begin{cases}\dfrac{\lambda J_1\left(\frac{2\pi\Delta_1}{\lambda}\right)}{\Delta_1 r_c}, & \Delta_1\neq 0\\ \pi/r_c, & \Delta_1=0\end{cases}. \quad (19)$$

Since $\pi/r_c$ is the amplitude at (0, 0), (8) is obtained after normalization.

Similarly, it is assumed that $r_c\gg\Delta_2$ the electric field along the *x*-axis with $x=\Delta_2$ can be derived a closed form in the same manner as for $E_x(\Delta_1,0)$, resulting in

$$E_x(0,\Delta_2)\approx\begin{cases}\dfrac{\lambda J_1\left(\frac{2\pi\Delta_2}{\lambda}\right)-2\pi\Delta_2 J_2\left(\frac{2\pi\Delta_2}{\lambda}\right)}{\Delta_2 r_c}, & \Delta_2=0\\ \pi/r_c, & \Delta_2=0\end{cases} \quad (20)$$

After normalization, (9) is obtained.

## Acknowledgement

This work was supported by a Project Grant in Excellence Center at Linköping-Lund in Information Technology (ELLIIT) Call D.